\documentstyle[11pt,aaspp4,psfig]{article}
\begin{document}

\title{On the Number of Comets Around White Dwarf Stars:  Orbit Survival During the Late Stages of Stellar Evolution}
\author{Joel Parriott}
\affil{Department of Astronomy, University of Michigan \\ Ann Arbor, MI  48109 \\ joelp@astro.lsa.umich.edu}
\and
\author{Charles Alcock}
\affil{Institute of Geophysics and Planetary Physics, Lawrence Livermore National Laboratory \\ Livermore, CA  94550 \\ alcock@igpp.llnl.gov}

\begin{abstract}
The accretion of comets onto DA white dwarfs can produce observable metal 
absorption lines.  We show here that comet systems around the progenitor main
sequence star are vulnerable to being lost during asymptotic giant branch
mass loss, if the mass loss is sufficiently asymmetric to impart modest linear
momentum to the white dwarf.  This may have bearing on the frequency of 
observation of heavy elements in white dwarf stars, and on inferences regarding
the frequency of comet systems, if the imparted linear velocities of white
dwarfs can be estimated.
\end{abstract}


\section{Introduction}

The current model for solar system comets, based upon the classic theory
of Oort \markcite{comet1}(1950), contends that a vast cloud of cometary objects 
orbits the sun, and that this cloud consists of three components.  The 
first, most inner, component is
a disk-like structure of $\gtrsim 10^{10}$ comets referred to as the Kuiper belt
(Edgeworth \markcite{comet1.5}1949; Kuiper \markcite{comet2}1951).  
The Kuiper belt has been proposed as the source for the Jupiter-family 
short-period (SP) comets, and to lie about $40 - 10^3$ AU from the sun 
(see Weissman \markcite{comet4}1995, and 
Luu {\em et al.} \markcite{comet5}1997).  
The second component, referred to as the inner Oort cloud (or ``Hills'' 
Cloud; Hills \markcite{comet5.5}1981), is believed to be a large fatter disk
containing $10^{12} - 10^{13}$ objects lying $\sim 10^3 - 2 \times 10^4$ AU from
the sun.  The Hills Cloud is proposed as a source for both long-period (LP) and 
Halley-type SP comets (Levison \markcite{comet3}1996).
The third component, is the classical Oort cloud (Oort \markcite{comet1}1950).
It is a roughly spherical cloud of $10^{11} - 10^{12}$ cometary objects, has a
nearly isotropic velocity distribution, and extends from $2 \times 10^4$ 
to $2 \times 10^5$ AU.  This component, which is physically contiguous with 
the Hills Cloud, yet dynamically distinct from it, is postulated to be a 
reservoir of LP comets which are brought into the inner solar system by 
perturbations due to passing stars, molecular clouds, and the galactic tidal 
field.  It is likely that comets from the more populous Hills cloud region 
are continuously replenishing the randomized orbits of the outer Oort cloud,
which are more easily lost from the stellar system due to their 
susceptibility to perturbation.  While the formation history of the outer Oort
cloud is disputed, the inner two components are thought to have formed out of
the proto-planetary disk.  For a further discussion of these topics and more, 
see these excellent reviews and references therein 
(Duncan and Quinn \markcite{comet7}1993; Weissman 
\markcite{comet8}1996; Levison \markcite{comet3}1996). 
This paper concerns the cloud components responsible for the LP comets, 
since they are the objects most likely to remain bound to the stellar 
system for the entire evolution of the central star.

It is reasonable to suppose that similar cometary systems exist around
other stars given quite particular formation conditions 
(Tremaine \markcite{es2}1993; 
Shull and Stern \markcite{es3}1995); 
however, the cold, diffuse nature of such an Oort-type cloud would make
it very difficult to directly detect
(Stern, Shull, and Brandt \markcite{subl}1990;
Stern, Stocke, and Weissman \markcite{es4.5}1991).
Dust shells (eg. $\alpha$ Lyr) and disk-like structures (eg. $\beta$ Pic), 
have been observed, via infrared excess, around $\lesssim 10\%$ of nearby stars 
(Aumann {\em et al.} \markcite{es6}1984; 
Aumann \markcite{es6.5}1985;
Walker and Wolstencroft \markcite{es7}1989; 
Aumann and Probst \markcite{es7.5}1991;
Oudmaijer {\em et al.} \markcite{es8}1992; 
Zuckerman \markcite{es9}1993;
Lagrange-Henri \markcite{es10}1995;
Plets {\em et al.} \markcite{es10.5}1997).
These shells and disks may contain cometary material and planets, but 
definitive proof of the presence of extra-solar comets is elusive  
(Weissman \markcite{es11}1984; 
Harper, Lowenstein, and Davidson \markcite{es12}1984;
Matese, Whitmore, and Reynolds \markcite{es13}1989; 
Zuckerman and Becklin \markcite{es13.5}1993;
Beust {\em et al.} \markcite{es14}1994;
Hubeny and Heap \markcite{es14.5}1996;
Grady {\em et al.} \markcite{es15}1997).
The direct detection of 
Hills Cloud and Kuiper-type cometary systems around nearby stars is made 
more promising by the rapid evolution of technologies such as improved infrared
detectors, new large infrared-optimized ground-based telescopes, adaptive 
optics, high-altitude and space-based infrared telescopes, and 
(sub-)millimeter arrays (Cruikshank, Werner, and Backman \markcite{es15.5}1994;
Burrows {\em et al.} \markcite{es16}1995; 
Roddier \markcite{es17}1995; Shao \markcite{es18}1996; 
Holland {\em et al.} \markcite{es19}1997).  While the detection
of these inner cometary systems, coinciding with large planets, may 
indirectly point to the presence of an Oort-type cloud, the null direct 
detections of a large outer cloud may continue.
Given such an apparent impasse for direct observation, this work concerns
possible implications for the indirect detection of Oort-type comet clouds 
around other single intermediate mass stars.  These stars should end up as 
white dwarfs during the normal stellar evolutionary sequence.

Alcock, Fristrom, and Siegelman \markcite{afs1}(1986, hereafter AFS) proposed 
that the 
accretion of a massive comet (e.g. $\sim 10^{17}$ g) onto a white dwarf star
could raise the heavy element abundances above detection limits in the
nearly pure hydrogen spectrum of a DA white dwarf.  
Most white dwarfs have hydrogen-rich atmospheres (DA) or helium-rich 
atmospheres (DB), which show no evidence for, or very low abundances of, 
elements heavier than hydrogen or helium.  
(For a good review of the optical characteristics and 
classification system of white dwarfs, see Wesemael {\em et al.} 
\markcite{wd1}(1993).  The presence of 
metals, yet very little hydrogen, in DB stars is not well understood, and so 
we concern ourselves with the relatively simple DA/DAZ stars.)
The dearth of detectable metals in DA and DAZ white dwarfs is believed to be 
caused by gravity-driven diffusive sedimentation of the heavier elements down 
out of the atmosphere (Alcock and Illarionov \markcite{wd2}1980a;
Vauclair, Vauclair, and Greenstein \markcite{wd3}1979; 
Fontaine and Michaud \markcite{wd4}1979).
According to the updated calculations of Paquette {\em et al.} 
(\markcite{wd5}1986), the time scale for this metal diffusion process is on the 
order of $10^4$ yr for a cool (${\rm T}_{\rm eff} \sim 10,000$ K) DA white dwarf
appropriate to the AFS analysis.
Despite the rare appearance of metal lines in cool DA white dwarfs, several
groups have made recent successful detections.  The following papers contain
metal abundances for DA/DAZ stars which are cool enough 
(${\rm T}_{\rm eff} \lesssim 12,000$ K) to be of interest in this work,
because their diffusion time scales are comparable to the comet accretion
time scale:
Dupuis, Fontaine, and Wesemael \markcite{wd9}(1993);
Shipman {\em et al.} \markcite{wd10}(1995);
Bergeron, Ruiz, and Leggett \markcite{wd8}1997;
Koester, Provencal, and Shipman \markcite{wd11}(1997);
and Bill\`{e}res {\em et al.} \markcite{wd12}(1997).  

The signature of comets around white dwarfs depends on two properties of
the intermediate temperature DA and DAZ stars.  The first is the short
sedimentation time, which implies that any heavy elements seen in the
atmosphere are not primordial and must have been added recently.
The second property of white dwarfs invoked by AFS is the 
small mass ($\lesssim 10^{23}$ g) 
of the atmosphere and outer convective envelope of these white dwarfs.  Given
their small atmospheres, the mass of material accreted need not be very large
in order to produce detectable metal lines within the remarkably pure hydrogen 
spectrum.  The accretion of one large comet can add enough material to be 
observed in this manner.

It should be pointed out that interstellar grain accretion is the standard 
explanation for the presence of calcium and magnesium lines in cool white dwarfs.
The most recent work on the accretion-diffusion model for white dwarfs can be 
found in Dupuis {\em et al.} \markcite{wd7}(1992, 1993).  Major objections to
this model have been raised by Aannestad {\em et al.} \markcite{wd13}(1993), 
who cannot
account for observed absolute calcium abundances given the low 
density interstellar medium around the solar neighborhood white dwarfs.  
They calculate limits on calcium accretion rates that are several orders of 
magnitude too low for observed abundances 
(cf. Alcock and Illarionov \markcite{afs3}1980b).

AFS point to the detection of calcium in the cool DAZ white
dwarf G74~-~7 (${\rm T}_{\rm eff} \sim 7000$ K; Lacombe {\em et al.} 
\markcite{afs2}1983) as an
example of a possible recent accretion of a circumstellar comet, rather than 
the presumably negligible accretion of interstellar material (Alcock and
Illarionov \markcite{afs3}1980b).  
(A very recent abundance determination for G74~-~7 can be found in 
Bill\`{e}res {\em et al.} \markcite{wd12}(1997).)
Using Joss's \markcite{afs4}(1974) estimate of the rate of accretion
of such a massive comet onto the Sun of $10^{-2} {\rm yr}^{-1}$, and the
observed flat distribution of comet orbits in periastron distance (Fernandez
\markcite{afs5}1982), AFS
estimated that the rate of comet accretion onto a white dwarf (with radius 
$\sim~0.01 {\rm R}_{\odot}$) surrounded by an Oort-type cloud and planetary
system similar to that of the solar system should be $\sim~10^{-4} 
{\rm yr}^{-1}$.  
Since the estimated time between accretions is
of the same order as the diffusion time for a cool DA star, one would 
expect that a white dwarf with an Oort-type cloud should maintain detectable 
metal lines for a significant fraction
of its lifetime.  Given the metal abundances, one can place limits on the 
accretion rate from such a cloud and therefore define limits on the size
of the cometary system surrounding the white dwarf.

AFS investigated the survival of a LP comet cloud population around
a white dwarf during the periods of vigorous mass loss and very high 
luminosity through the post-asymptotic giant branch and 
pre-planetary nebula phases. 
Short-period comets at Kuiper belt distances from the star would 
be destroyed during this evolution (Stern, Shull, and Brandt 
\markcite{subl}1990).  
AFS integrated orbits of LP comets (with a distribution similar 
to that inferred for the outer Oort cloud)
in a stellar system which undergoes a gradual, spherically symmetric mass loss,
with the mass of the central star given as a function of time, $M(t)$.  They
found that even when a significant fraction a star's
mass is lost during the pre-white dwarf phase of its evolution more than half
of the comets in high-eccentricity, LP orbits remain bound, providing
a supply of star-grazing comets which can be accreted by the star.  The main
physical reason for this result is that these comets spend most of their time
at apastron, where they are more insensitive to changes in the potential energy
of the system due to their low kinetic energy, and where they are out of danger
of evaporation by the central star during its highly luminous red giant phase.
The semi-major axes of the orbits increase due to the decreasing depth of the
potential well of the system, but they do remain bound.

An important assumption made by AFS was that the mass loss during the 
advanced stages of stellar evolution is spherically symmetric.  
However, even a modest recoil velocity ($\sim 100$ m/s) of the central star 
due to an asymmetry in the mass loss can be comparable to the speed of a comet 
near apastron, and as such should not be neglected when studying the 
survivability of such comets.  Numerous observational studies of late-type 
stars and planetary nebulae have found that the majority of these objects 
show a dramatic departure from spherical symmetry
(eg. Zuckerman and Aller \markcite{ml0}1986;
Balick \markcite{ml0.5}1987;
Bujarrabal, Alcolea, and Planesas \markcite{ml1}1992;
Plez and Lambert \markcite{ml2}1994;
Kastner {\em et al.} \markcite{ml3}1996;
Manchado {\em et al.} \markcite{ml3.2}1996).  
For some recent HST results, see
Sahai {\em et al.} \markcite{ml3.4}(1997), 
Trammell and Goodrich \markcite{ml3.6}(1997), and
Sahai {\em et al.} \markcite{ml3.8}(1998).
While most planetary nebulae are aspherical, they typically have a high 
degree of other forms of intrinsic symmetry, such as elliptical, bipolar, and
point-symmetry (see a review by Pottasch \markcite{ml4}1995, 
and references therein; 
Manchado, Stanghellini, and Guerrero \markcite{ml5}1996; 
G\'{o}rny, Stasi\'{n}ksa, and Tylenda \markcite{ml6}1997).  There are also a
small number of objects which have no clear morphological symmetry, and are 
classified as ``irregular.''
The formation and evolution conditions that lead from relatively spherical
red giants to the diverse morphologies of planetary nebulae is a matter of 
active research, and beyond the scope of this paper.
We are interested in irregularities and departures from perfect symmetry, 
and the bipolar proto-planetary nebula 
OH 231.8+4.2 is an extreme case of such an imperfection.  It has been 
shown that while OH 231.8+4.2 is axially symmetric, its two lobes 
differ {\em intrinsically} from one another 
(eg. Kastner {\em et al.} \markcite{ml7}1992;
Alcolea, Bujarrabal, and S\'{a}nchez Contreras \markcite{ml8}1996).
Alcolea, Bujarrabal, and S\'{a}nchez Contreras \markcite{ml8}(1996) find that 
the southern
lobe is $20 \%$ more massive than the northern lobe.  Such a significant
departure from symmetry is rare, but only small differences are required 
to impart the recoil velocities of interest in this paper.  The model fits 
to apparently symmetric morphologies are not absolutely perfect, and it is 
conceivable that small intrinsic asymmetries, at levels well below those 
seen in OH 231.8+4.2, have been going undetected in other objects.

In this work we extend the calculation of AFS to account for a simple 
asymmetric mass loss.  Such a mechanism imparts a net impulsive linear 
momentum to the 
central star, which, as mentioned above, is important once its magnitude
becomes comparable to the speed of a comet at apastron.  We investigate the
interesting parameter space of the problem,
using Monte Carlo simulations, to identify when the mass loss rate, and thus
the impulse velocity of the star, becomes large enough to significantly reduce
the number of comets able to survive the mass loss.

In $\S$ 2 we present the calculation used to model the effects of asymmetric
mass loss on an initial distribution of comet orbits similar to that of the
Oort-cloud.  The results of this calculation, in terms of
the survivability of these comet orbits given a wide set of parameters that
we feel covers the interesting parameter space of the problem, are presented
in $\S$ 3.  
Finally, in $\S$ 4 we discuss the implications of these calculations for 
researchers
in the field interested in the probability that a population of star-grazing
comets could survive an asymmetric mass loss, given a knowledge of the
conditions of the post asymptotic giant branch stage of a given star's
evolution.

\section{Modeling the Asymmetric Mass Loss and Comet Ejection}

We are interested in investigating the implications for an initial 
distribution of comet orbits, similar to those one would expect from an
Oort-type cloud, when the central star of the system undergoes
a smooth asymmetric mass loss.  The asymmetric mass loss manifests itself 
by providing a net impulsive rectilinear acceleration to the central potential 
of the system.  Of particular interest is the magnitude of this impulsive force
necessary to unbind even the high-eccentricity long-period
comet orbits, which proved to have a high survival rate for the spherically
symmetric mass loss in AFS.

We treat the rectilinear acceleration due to the asymmetric mass loss as
an additive fictitious force per unit mass ${\bf f}_{r}$
in the equation of motion for the comet. The magnitude of this fictitious 
force is proportional to the mass loss rate,
which one would qualitatively expect, and to the total velocity change of
the star during the mass loss phase.  For simplicity, the magnitude and 
direction of the mass loss are taken to be constant for the entire mass loss 
interval.  Since the
mass loss is not restricted to the initial orbital plane, angular momentum is 
not conserved, and the motion in the direction perpendicular to the plane must
be included.  The vector form of the equation of motion, realized in the 
instantaneous rest frame of the star, becomes
\begin{equation}
\label{eq:motion}
\frac{d^{2} {\bf r}}{dt^{2}} = - \frac{GM(t)\: {\bf r}}{r^{3}} - {\bf f}_{r},
 \;\;{\rm where}\;\;\; {\bf f}_{r} = {\bf \Delta v}\: \frac{\stackrel 
{\displaystyle .}{M}\!\!(t)}{M_{{\rm f}} - M_{{\rm i}}}\;.
\end{equation}
${\bf \Delta v}$ represents the total velocity change imparted to the star, 
due to this rectilinear acceleration, during the mass loss phase.
For $M(t)$ we assume the smooth analytic mass loss function used by AFS, which 
has the form
\begin{equation}
\label{eq:mlf}
M(t) = \frac{1}{2}(M_{{\rm i}} + M_{{\rm f}}) - \frac{1}{2}(M_{{\rm i}} 
- M_{{\rm f}})\:{\rm tanh}\: \frac{2t}{T}\;,
\end{equation}
where $M_{{\rm i}}$ is the initial mass,
$M_{{\rm f}}$ is the final mass, and $T$ is the mass loss interval.  We
concern ourselves with the time interval $-3T < t < 3T$, so that the time of
vigorous mass loss is included fully and the error 
in the initial and final masses falls below one part in $10^4$.  
$\stackrel{\displaystyle .}{M}\!\!(t)$ is then obtained by differentiation of
equation (\ref{eq:mlf}).

We choose to parameterize the total impulsive velocity of the star 
${\bf \Delta v}$ in terms of the ratio of the 
semi-major axis of the orbit $a$ to the period $P$ multiplied by a 
dimensionless factor $\Lambda$:
\begin{equation}
\label{eq:vel}
{\bf \Delta v} =  \Lambda \frac{a}{P} {\bf \hat{n}},
\end{equation}
where ${\bf \hat{n}}$ is a unit vector in the direction of the rectilinear
acceleration (opposite the direction of the mass loss).
This is a convenient scale because $a/P$ is of the same order as the 
characteristic velocity of the problem, i.e. the average velocity of a 
comet around its orbit.  An impulsive velocity imparted to the star which
is much larger than this characteristic velocity would be expected simply to 
separate the star from the slowly moving cometary system, and a much smaller
velocity would have negligible influence on the state of stellar system.  The
interesting effects for this occur when $\Lambda \sim 1$. 

The dynamical equations (\ref{eq:motion}) are invariant under the 
transformation:
\begin{eqnarray}
M & \longrightarrow & \gamma M  \nonumber \\
{\bf r} & \longrightarrow & \xi {\bf r}  \nonumber \\
t & \longrightarrow & \xi^{\frac{3}{2}} \gamma^{-\frac{1}{2}} \;t \\
{\bf \Delta {\rm v}} & \longrightarrow & \xi^{-\frac{1}{2}} \gamma^{\frac{1}{2}}\; {\bf \Delta {\rm v}}  \nonumber \\
\stackrel {\displaystyle .}{M}\!\!(t) & \longrightarrow & \xi^{-\frac{3}{2}} \gamma^{\frac{3}{2}} \stackrel {\displaystyle .}{M}\!\!(t) \: . \nonumber
\end{eqnarray}

Units for the computation were selected such that the initial $a = GM = 1$ 
for each orbit integration, and so the initial period of the orbit $P = 2\pi$.  
The initial conditions of each comet orbit are then completely determined
by the eccentricity squared $e^{2}$, and the starting phase $w_{3}$ (the
fractional area of the ellipse since the last periastron; Goldstein 
\markcite{cm}1981). 
 
If one starts with a particular choice of a physical problem ($a_{\rm i}, 
\Delta {\rm v}, M_{\rm f}, M_{\rm i}, 
{\stackrel{\displaystyle .}{M}}_{\rm max}$), the transformation to the 
units used in the code, as given by this formalism, is
\begin{eqnarray}
\gamma &  = & {(GM_{\rm i})}^{-1}, \\
\xi & = & {a_{\rm i}}^{-1}\: ;
\end{eqnarray}
therefore,
\begin{equation}
\label{eq:convert}
\Lambda  =  \Delta {\rm v} \frac {\stackrel {\displaystyle .}{M}_{\rm max}}{M_{{\rm i}} - M_{{\rm f}}} \; \frac {a_{\rm i}^2}{GM_{\rm i}},
\end{equation}
where $\stackrel {\displaystyle .}{M}_{\rm max}$ is the maximum value of the
mass loss rate (functionally, the value of $\stackrel {\displaystyle .}
{M}\!\!(t)$ at $t = 0$).  
For example, one obtains $\Lambda \approx 6.7$ for the following set of
typical parameters: $a_{\rm i} = 10^{4} AU, \Delta {\rm v} =
300 {\rm m/s}, M_{\rm i} = 2 M_{\odot}, M_{\rm f} = 0.8 M_{\odot},
{\stackrel{\displaystyle .}{M}}_{\rm max} = 10^{-4} M_{\odot}/{\rm yr}$.  The
calculation of the fraction of comet orbits which remain bound after the
mass loss, $B$, versus the ratio of the initial period to the mass loss 
interval,
$P/T$, corresponding to these parameters can be found by interpolating 
between the second and third column plots of Figure~4, given an orbit 
eccentricity.  We address these plots fully in Section~III.

In order best to
simulate the physical nature of the problem, Monte Carlo techniques
were used to provide a random initial $w_{3}$ and mass loss direction for 
each integration.  Since a random direction for the mass loss is
used, the inclination of the initial orbital plane is removed from the
problem.  To $\sqrt{N}$ accuracy, the distribution of 
initial comet orbits is uniform with respect to $w_{3}$ and the mass loss 
direction on the sphere, where $N = 10,000$ for all of the calculations
shown in this paper.

The equation of motion of an orbit was integrated as a function
of time in Cartesian coordinates, using an adaptive time step,
fourth order Runge-Kutta method.  The code was tested by 
integrating Kepler orbits, with
the mass of the star remaining fixed, and further by reproducing the 
results of AFS in the spherical mass loss case ($\Lambda = 0$). 

\section{Results}

We begin by obtaining the value of the energy of a comet at the end of each
orbit integration, $E_{\rm f}$, as a function of the initial phase, $w_{3}$, 
for each 
parameter set: $e^{2}$, ${\bf \Delta v}$, $P/T$, and $M_{{\rm f}}/ M_{{\rm i}}$.
The fraction of orbital initial phase with $E_{\rm f} > 0$ represents the 
fraction of comets which have become unbound and escaped from the system.  
For each set of integrations we determine the fraction of comets that
remain bound after mass loss, which we will denote by $B$. 
The values of $E_{\rm f}$ versus $w_{3}$ lie on a single smooth curve for the 
symmetric mass loss case (${\bf \Delta v} = 0$; see AFS); 
however, the final energy curves become much more complex once the
asymmetric mass loss is introduced.  

Figure~1 contains representative plots of this $E_{\rm f}$ versus $w_{3}$
relationship for a specific set of parameters.  The calculation in Figure~1(a) 
corresponds to the case where the direction of the mass loss (opposite the
direction of the rectilinear acceleration) is restricted to the direction of
initial apastron in the initial orbital plane.  Roughly $2/3$ of the 
orbits in this case have become unbound ($E_{\rm f} > 0$), and the bound orbits
are found toward the middle range of $w_{3}$.  The curve is quite smooth is the
region $0.35 < w_{3} < 0.85$, but spikes appear outside this range.
In Figure~1(b), we
incorporate the entire randomness of both direction and initial phase for the
same set of parameters as in Figure~1(a).  Note the filling of the $E_{\rm f} 
- w_{3}$ 
parameter space bounded by seemingly smooth curves.  Closer observation
also reveals two bounding curves at work on the top edge of the region
for $w_{3} < 0.5$.  The extrema of $E_{\rm f}$ as well as the location of 
the global minimum are the same as in Figure~1(a).  There are also a greater
fraction of bound orbits in this case than for the restricted direction case.
The full Monte Carlo 
realization of the calculation introduces an entirely new level of complexity 
and richness to the problem.

\vspace{0.25in}

\centerline{\hbox{
\psfig{figure=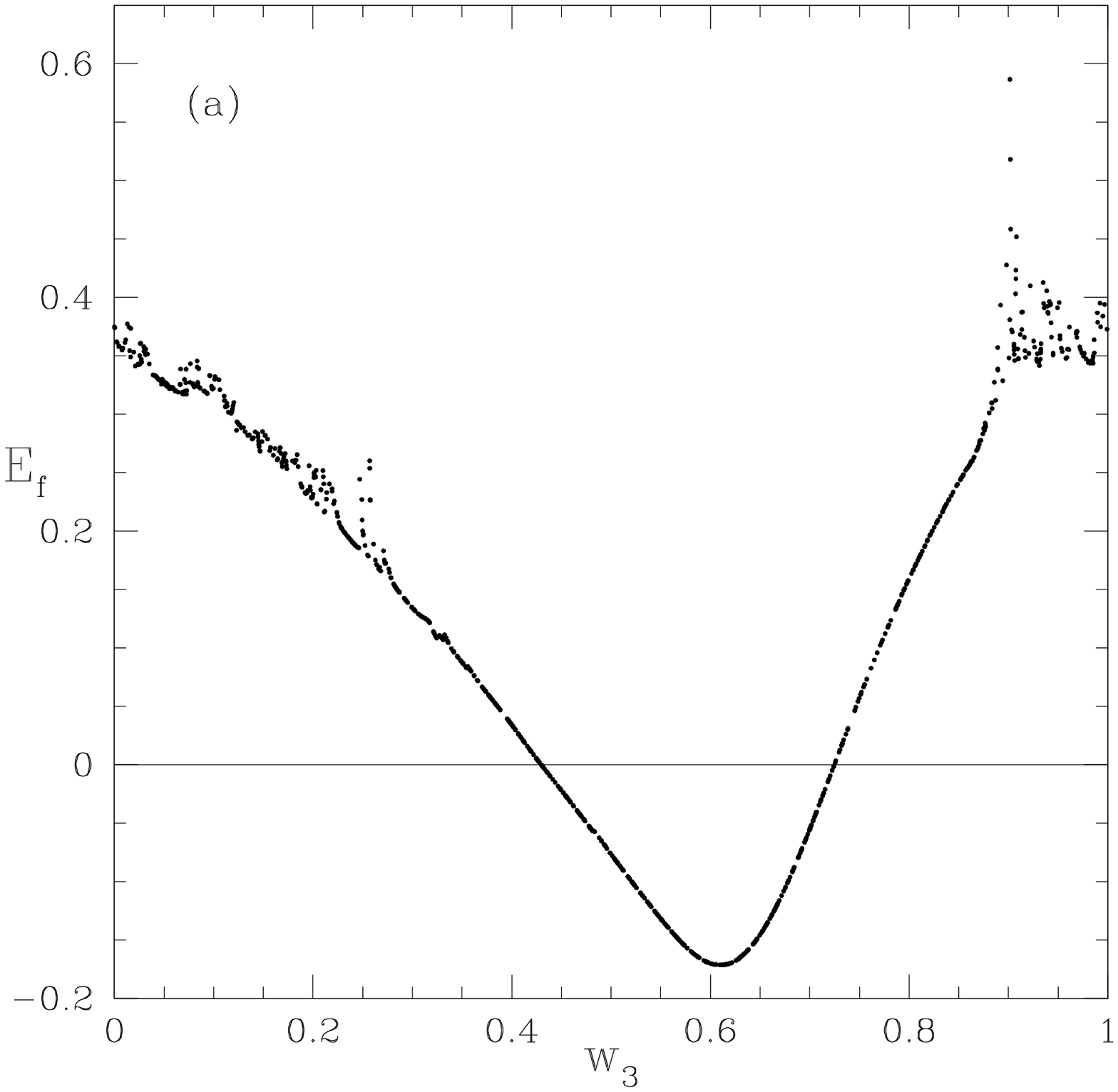,height=3in,width=3in}
\psfig{figure=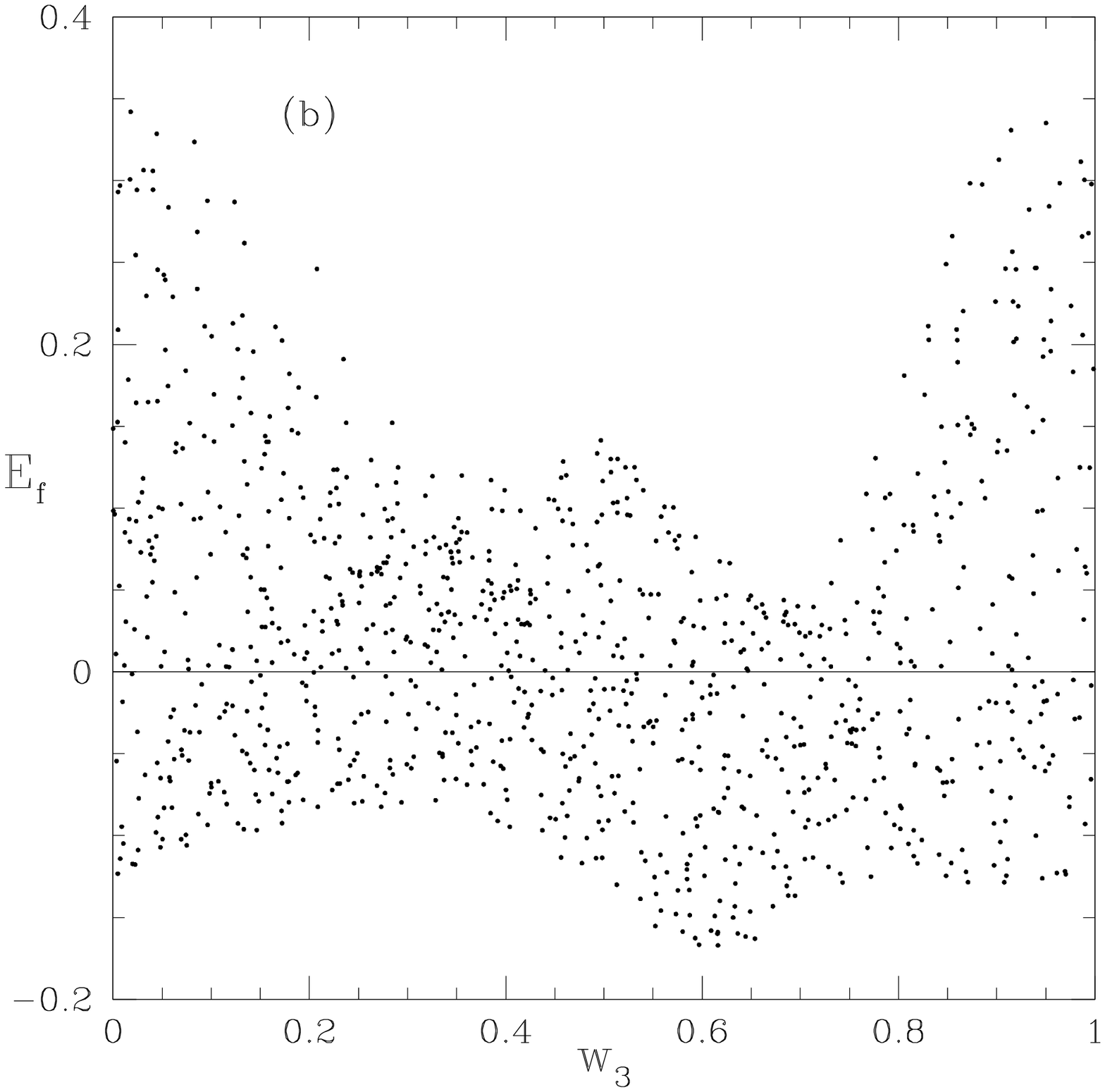,height=3in,width=3in}
}}

\figcaption[f1a.eps,f1b.eps] {
The final energy of an orbit as determined by the initial phase
$w_{3}$ (the fraction of the area of the orbital ellipse from periastron at the
start of the integration).  Since $GM = a = 1$, the initial energy of
all comets is $E = -0.5$.
(a) Mass loss restricted to the direction of initial apastron for all $w_3$.
Squared eccentricity $e^{2} = 0.8 $, multiplicative factor of the
mass loss $\Lambda = 4$, mass ratio $M_{{\rm f}}/M_{{\rm i}} =  0.4$, and
period to mass loss interval ratio $P/T = 0.9$.
(b) Fully random simulation (both $w_{3}$ and direction) for the same parameters
as in (a). Note change in abscissa scale.
\label{f1}}

\vspace{0.25in}

We repeat these calculations for different values of $P/T$ in order to sample
different periods in terms of the mass loss interval.  In Figures~2-6 we show 
mosaics of plots of 
the bound fraction of comets $B$ as a 
function of the initial period to mass loss interval ratio $P/T$, for different 
mass loss ratios.  The bound fraction obtained from each plot similar to
that in Figure~1(b) then represents a single point in these figures.
The values lie on a smooth curve, and the $\sqrt{N}$ fluctuations are
approximately the same size as the dots on the plot.
As was the case for symmetric mass loss,
small values of $P/T$ approximate the adiabatic limit, where the mass
loss takes place very slowly.  Large values of $P/T$ represent the sudden
mass loss approximation (Hills \markcite{mlf}1983), and the points
asymptotically
approach the line representing this limit, as expected.  In order to verify that
the limit is truly achieved in every case, we carried out full calculations
increasing $P/T$ up to a value of $2000$ and found each asymptote to be correct.
In the energetically susceptible orbit region around $P/T \approx 1$ we
see a minimum in the curve, as in the case for symmetric mass loss, but the
additional structure present in that region is quite interesting, yet
unexplained physically.  Even considering the
random fluctuations introduced by the Monte Carlo techniques, the structure

\centerline{\psfig{figure=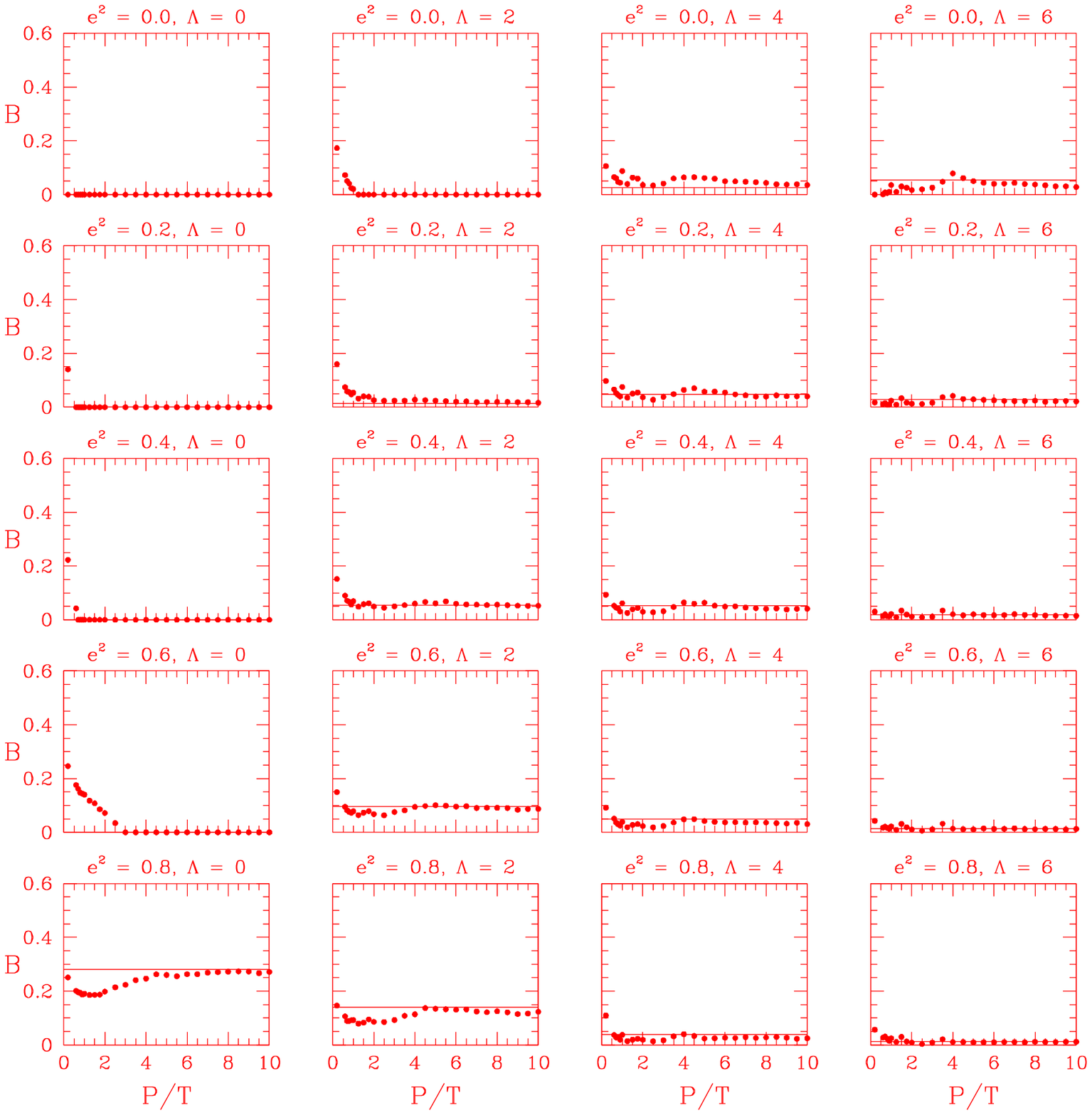,height=7.25in,width=7.125in}}
\figcaption[f2.eps] {
The probability $B$ that a comet remains bound
as a function of the ratio of initial period $P$ to mass loss interval $T$.
This mosaic represents a large parameter space range of initial squared
eccentricities $e^2$ and multiplicative factors for the mass loss $\Lambda$,
for a mass ratio $M_{{\rm f}}/M_{{\rm i}} =  0.1$.  The
$\sqrt{N}$ random errors are represented by the size of the points, and
the same error exists for the line.
\label{f2}}

\centerline{\psfig{figure=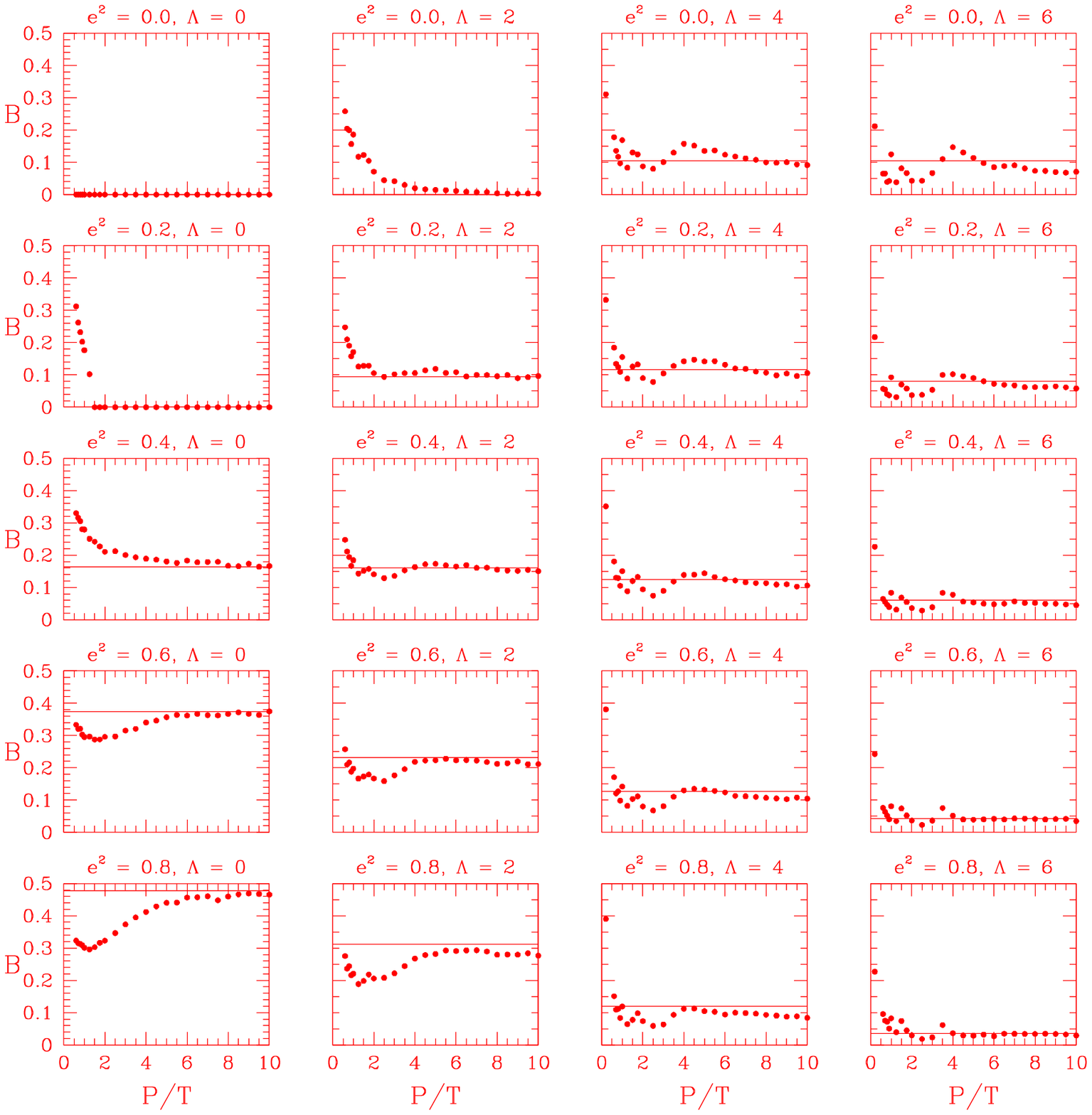,height=7.25in,width=7.125in}}
\figcaption[f3.eps] {
A mosaic similar to that in figure~2, for a mass ratio
$M_{{\rm f}}/M_{{\rm i}} = 0.2$. Note change in abscissa scale.
\label{f3}}

\centerline{\psfig{figure=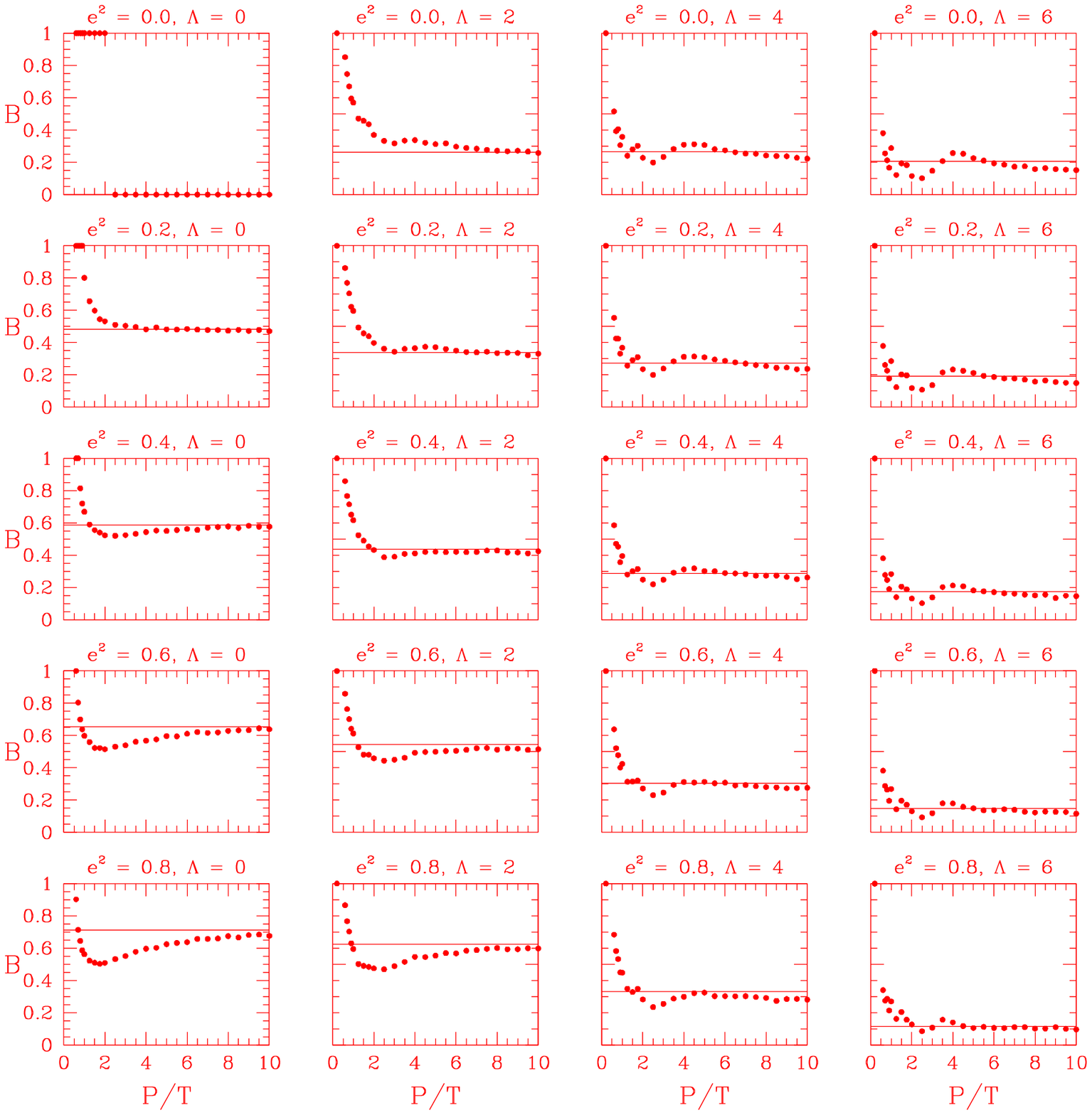,height=7.25in,width=7.125in}}
\figcaption[f4.eps] {
A mosaic similar to that in figure~2, for a mass ratio
$M_{{\rm f}}/M_{{\rm i}} = 0.4$. Note change in abscissa scale.
\label{f4}}

\centerline{\psfig{figure=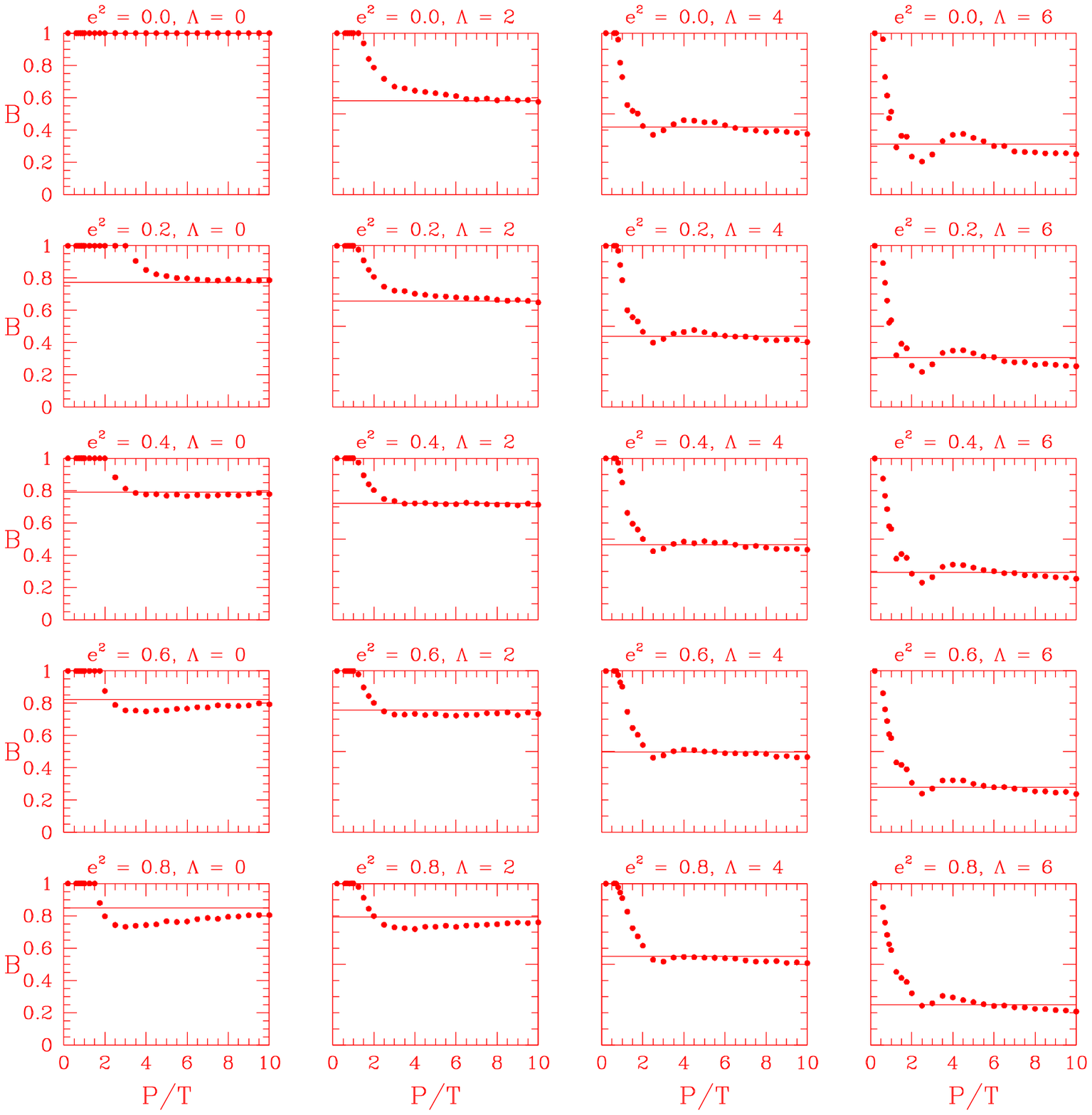,height=7.25in,width=7.125in}}
\figcaption[f5.eps] {
A mosaic similar to that in figure~2, for a mass ratio
$M_{{\rm f}}/M_{{\rm i}} = 0.6$. Note change in abscissa scale.
\label{f5}}

\centerline{\psfig{figure=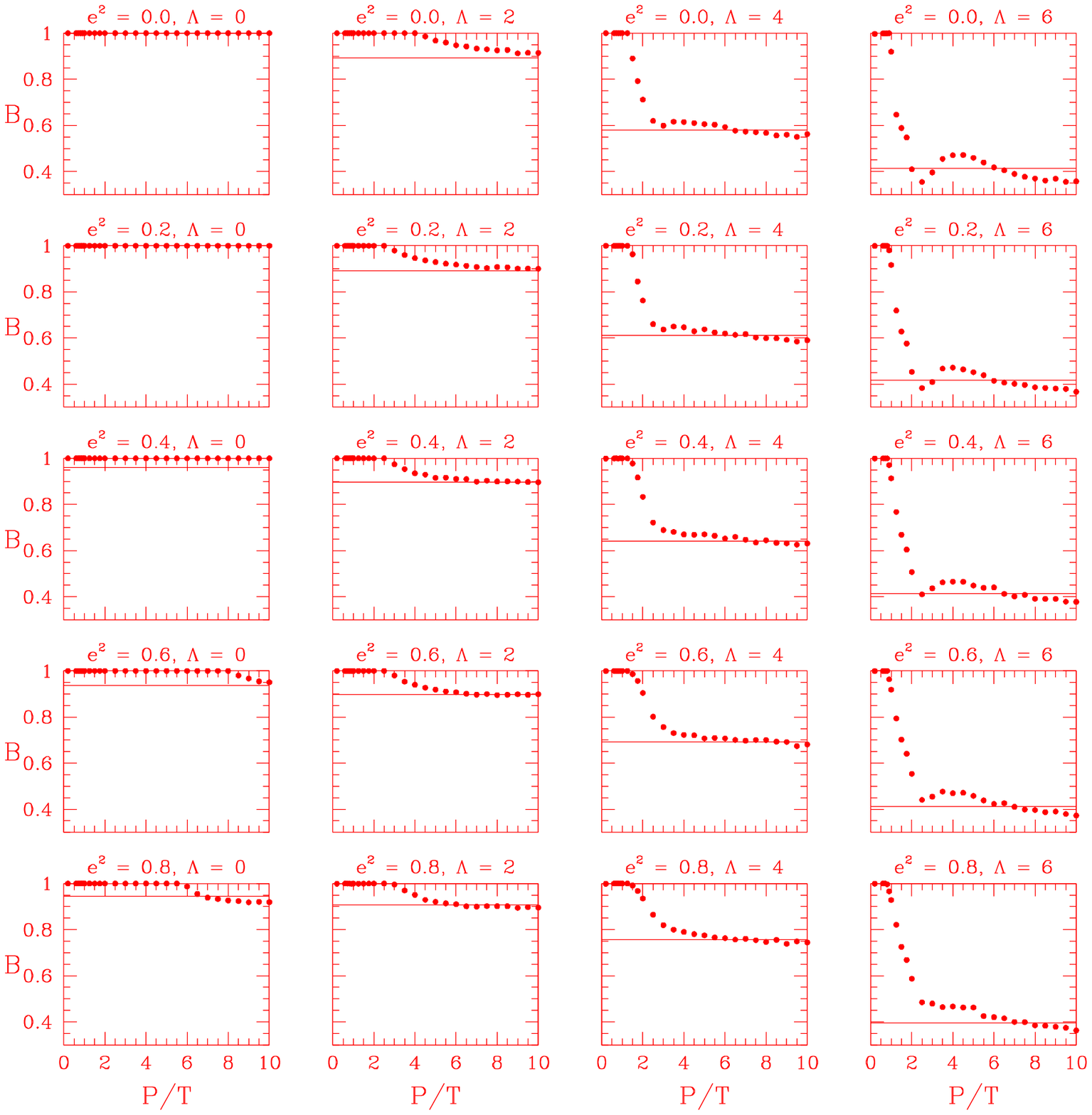,height=7.25in,width=7.125in}}
\figcaption[f6.eps] {
A mosaic similar to that in figure~2, for a mass ratio
$M_{{\rm f}}/M_{{\rm i}} = 0.8$. Note change in abscissa scale.
\label{f6}}

\clearpage

\noindent is real.  We believe that these mosaics provide a coverage of the
interesting parameter space relevant to this calculation, in terms of
the strength of the mass loss, mass loss ratios, and initial
eccentricities.  With such a representation of different scenarios,
one can see the cases where a significant number of comets will remain
in order to become probable future star-grazers for the white dwarf.  As was the
case for symmetric mass loss, it is precisely the highly eccentric
orbits which remain more bound as the magnitude of the rectilinear acceleration
increases.  The strength of the mass loss, given the scaling
of the mass loss amplitude used, need not become
very large in order to lose most of the comets.

The behavior of the asymptotic sudden mass loss limits also possesses
some interesting details.  In Figures~7 \& 8 we have plotted these asymptotes 
as a function of the multiplicative factor, $\Lambda$, for the range of 
eccentricities, $e^{2}$, shown in the mosaics, but only for 
$M_{\rm f}/M_{\rm i} = 0.2 \& 0.6$ respectively.  These two mass loss fractions 
represent two interesting
cases in terms of the observations of the masses white dwarfs in clusters
where a main sequence turn-off point can be found, thereby allowing an
estimate of the mass of the white dwarf progenitor to be found. The case of
$M_{\rm f}/M_{\rm i} = 0.2$ is believed to be about the maximum mass loss
observed; the case in point is the white dwarf Sirius-B, which has a mass
of $1.1 M_{\odot}$ and is in a cluster with a turn off mass of 
$6.0 M_{\odot}$.   A $M_{\rm f}/M_{\rm i} = 0.6$ is about the mass loss
fraction expected for $1.0 M_{\odot}$ stars.
Both cases show very smooth curves which for the most part have the bound
asymptote decreasing monotonically with increasing $\Lambda$.  Both also
show apparent convergent point(s), near $\Lambda = 4-5$, which have a spread 
in $B$ that is not significant within $\sqrt{N}$ errors.

An interesting feature of Figure~7 is the
maximum in asymptote at this convergent point for circular and nearly 
circular orbits, yet one would intuitively expect the sudden mass loss 
limiting value for the surviving comets monotonically to
decrease as the effects of asymmetric mass loss become stronger (i.e. 
increasing impulse velocity of the central star).  This can be explained
because a certain fraction of the comet orbits have initial velocities
in the same direction as the rectilinear acceleration, and as such they
can ``catch up'' to the star for small to medium values of this acceleration.
Once the the acceleration becomes large enough they can no longer catch up.

Also note that the 
ordering from most bound fraction to least, for the given eccentricities, flips 
order at the convergent point(s).  At lower $\Lambda$, the higher the 
eccentricity the larger the bound fraction, but at higher $\Lambda$, the 
lower the eccentricity, the larger the bound fraction.  The circular orbits 
in Figure~8 do not follow this general trend at small $\Lambda$; they have
a larger bound asymptotic value than some or all of the more elliptical orbits,
and the effect must be interpreted as genuine because it is not within 
random errors.  $M_{\rm f}/M_{\rm i} = 0.6$ is the intersection of much of
the characteristic phase space for this problem, and so effects such as this 
are not easily isolated.

\vspace{0.25in}

\centerline{\psfig{figure=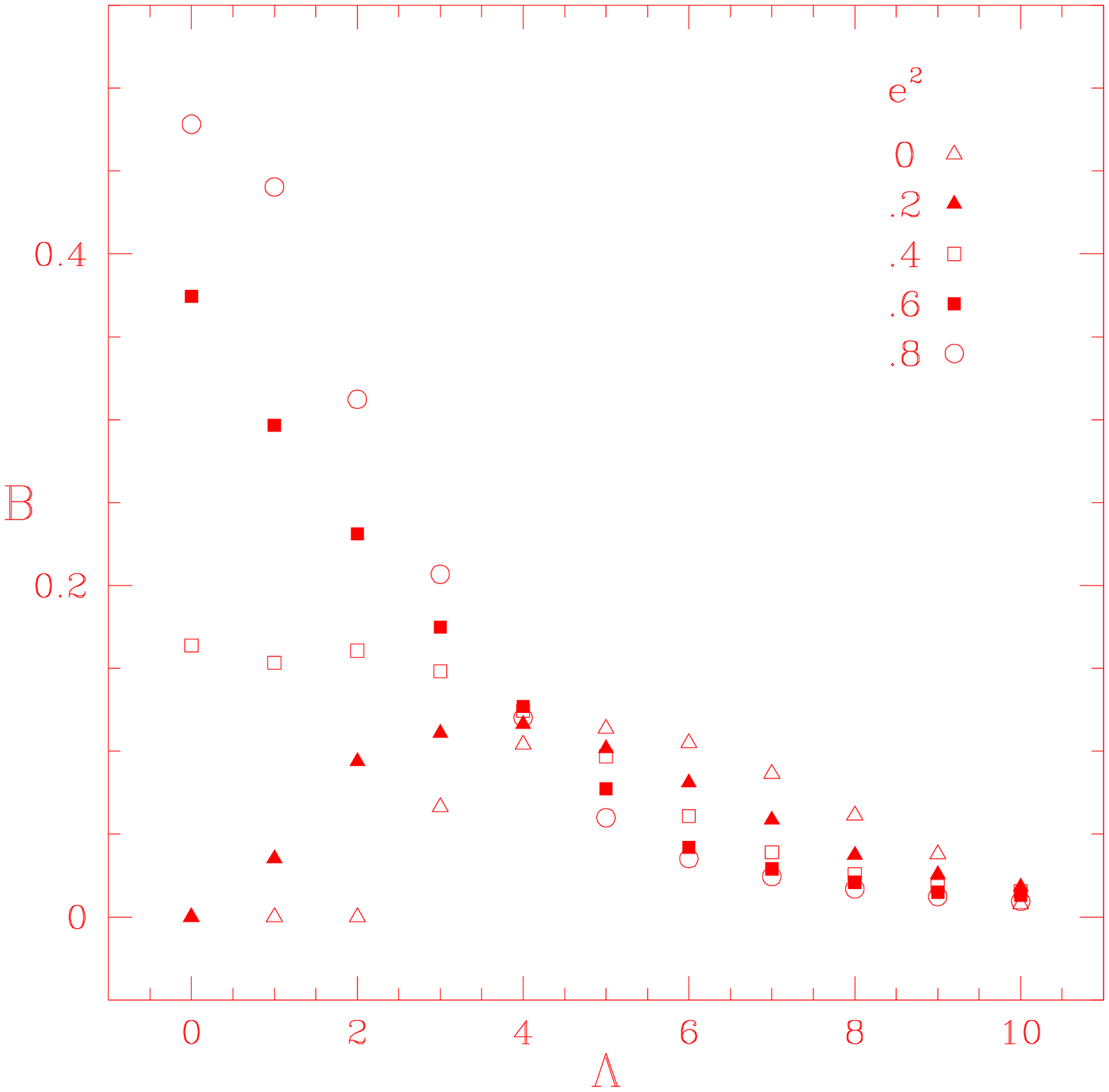,height=3in,width=3in}}
\figcaption[f7.eps] {
The value of $B$ for the sudden mass loss approximation asymptote
as a function of the mass loss multiplicative factor $\Lambda$, for the values
of $e^{2}$ represented in the mosaics, and for $M_{{\rm f}}/M_{{\rm i}} =
0.2$.
\label{f7}}

\vspace{0.25in}

\centerline{\psfig{figure=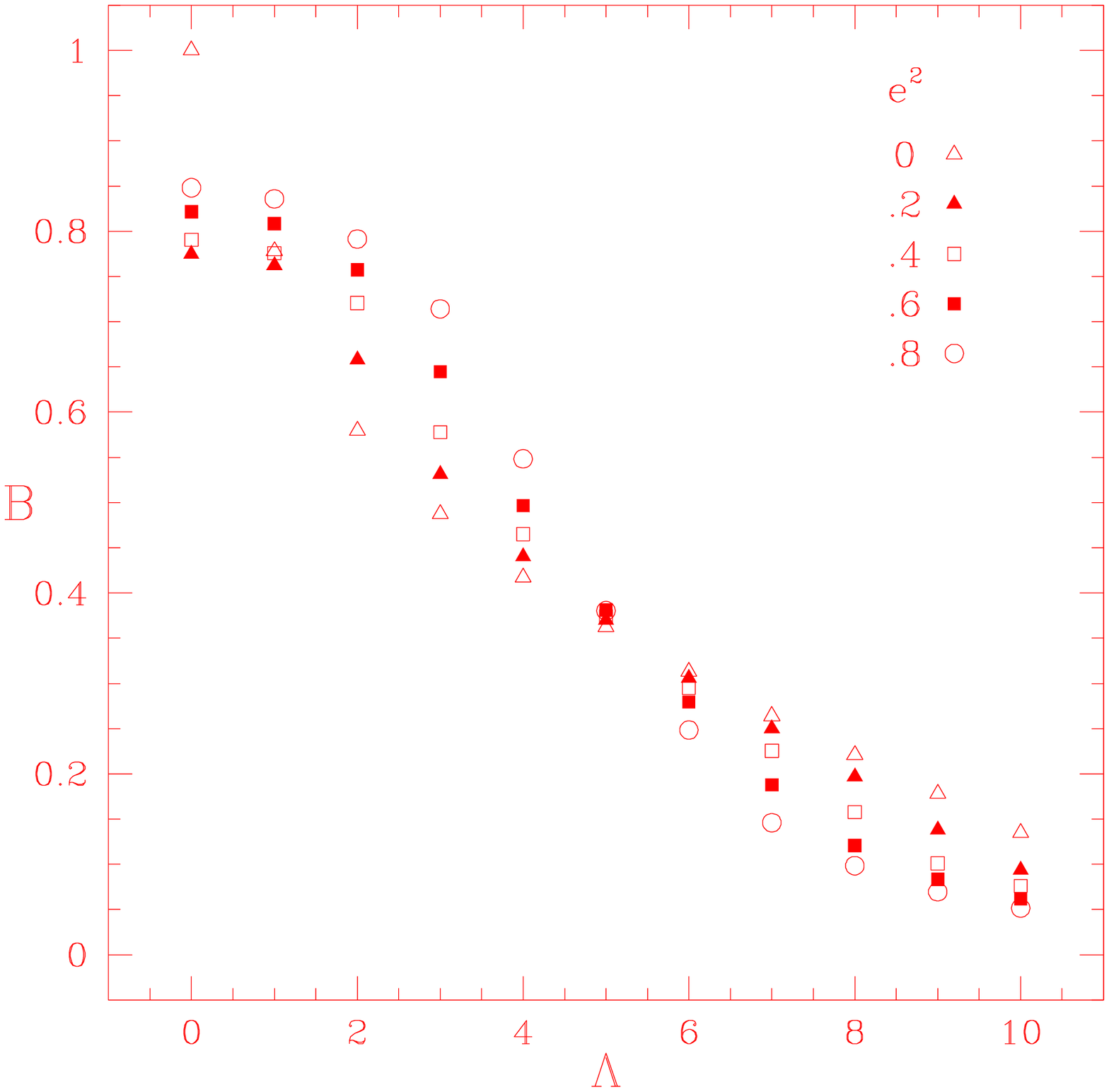,height=3in,width=3in}}
\figcaption[f8.eps] {
Similar to figure~7, for $M_{{\rm f}}/M_{{\rm i}}
 = 0.6$. Note change in abscissa scale.
\label{f8}}

\clearpage

\section{Discussion and Conclusions}

The results of these simulations will be useful to the reader interested in 
knowing the likelihood that a population of star-grazing long-period comets 
could have survived the mass loss stages of the white dwarf under study.
We have covered a large parameter space and provided a direct scaling 
relationship between our calculations and appropriate physical properties of 
a given physical system.  Comet survival can be estimated given
the comet's initial orbital eccentricity and semi-major axis and a knowledge
of the star's post asymptotic giant branch evolution:  progenitor initial mass,
maximum mass loss rate, net recoil velocity (if any) due to asymmetric mass 
loss, and final white dwarf mass.  We calculated an example given some
typical numbers for the above quantities and found a scaling that was well
within the simulation parameter space.

We have performed calculations in order to determine the effects of a
simple model for stellar asymmetric mass loss on an initial distribution of 
comet orbits for a system similar to the solar system.  
Asymmetric mass loss may explain the very small fraction ($ < 10 \%$) of 
white dwarfs with detectable metals when potential cometary systems 
are detected around a larger fraction ($ \sim 10 \%$) of nearby main sequence
stars, given the accretion signature theory of AFS.  Our calculations show 
that an impulsive velocity of the central star 
of only a few hundred m/s is enough to even drastically reduce the number of
long-period, star-grazing comets in an Oort-type cloud population.

\acknowledgments

CA wishes to thank Jim Liebert for a useful discussion.  We would
also like to thank Ben Zuckerman for his helpful comments.  JP was supported 
in part by the Computational Science Graduate Fellowship Program of the Office 
of Scientific Computing in the Department of Energy.
Research at LLNL is supported under Contract W-7405-ENG from the Department 
of Energy.

\end{document}